\documentclass[12pt]{article}
\usepackage{a4}
\usepackage{epsfig,url}
\usepackage{lineno}

\newcommand\uB{\underline{B}}
\newcommand\vp{\varphi}
\begin{document}
\title{Analytic Expressions\\ for Shielded Halbach Multipoles}
\author{Volker Ziemann \\ Thomas Jefferson National Accelerator Facility \\
  Newport News, VA, USA}
\date{February 26, 2026}
\maketitle
\begin{abstract}\noindent
We employ the method of images to derive analytic expressions for the
magnetic field of Halbach multipoles that are enclosed in high-permeability
shielding.
\end{abstract}
%
%
\section{Introduction}
Permanent magnets provide strong magnetic fields without using power supplies.
This makes them very popular in in today's energy-conscious
world striving for sustainability. So it is no surprise that they are used in
particle accelerators~\cite{FNAL,CBETA,SPRING,SLS2} as well. Many of the
multipole magnets evolve from Halbach's analytic expressions for two-dimensional
fields from~\cite{HALBACH}
which exploit the fact that the permeability of permanent magnet material is
very close to unity and that the fields of separate magnets are given by the
superposition of the fields from the individual blocks, as long as no ferromagnetic
material, such as iron, is close by.
\par
The absence of ferromagnetic material makes it impossible to account for
shielding the stray fields of the magnets with high-permeability materials.
In this report we generalize Halbach's analytic expressions to account for
magnetic shielding configurations that exhibit a high degree of symmetry.
This might prove useful in a first estimate of a magnet performance and before
resorting to numerical methods~\cite{OPERA,CHUBAR,SCHEER}.
\par
In the following sections we describe how to account for shielding with
infinite-permeability material by introducing image fields for magnetic
dipoles. We then apply
the theory to multipoles with a continuously rotating easy axis. In
Section~\ref{sec:seg} we calculate the expressions for segmented multipoles
followed by Section~\ref{sec:cube} on multipoles constructed from magnetic
cubes. Section~\ref{sec:conc} wraps up the report with the conclusions.  
\section{Image fields for magnetic dipoles}
%
\begin{figure}[tb]
\begin{center}
\includegraphics[width=0.7\textwidth]{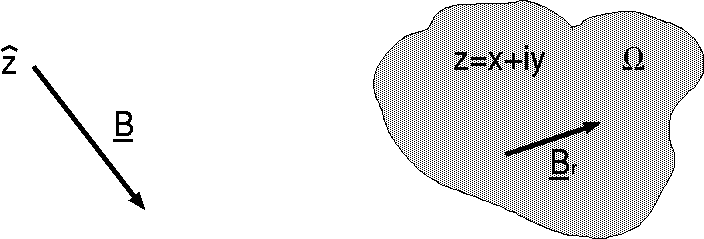}
\end{center}
\caption{\label{fig:geo}The geometry with the domain $\Omega$ containing permanent-magnet
material with remanent field $\uB_r$ and the field $\uB$ that it creates at position $\hat z$.}
\end{figure}
%
Halbach's derivations from~\cite{HALBACH} rely on the fact that Maxwell's
equations in two dimensions are equivalent to the Cauchy-Riemann equations~\cite{LR}
for one complex variable. All fields are therefore expressed as complex-valued
functions $\uB(z)=B_x+iB_y$, denoted by underscored symbols, of complex-valued positions
$z=x+iy$. In particular, the field generated by permanent magnets with remanent field
$\uB_r=(B_r)_x+i(B_r)_y$ can be written as~\cite{APB}
\begin{equation}\label{eq:Bdip}
  \uB^{\ast}(\hat z)=B_x-iB_y=\frac{1}{2\pi}\int_{\Omega}\frac{\uB_r}{(\hat z- z)^2}dx dy\ ,
\end{equation}
where $\Omega$ is the region that the permanent magnet occupies, $\uB^{\ast}$ is
the complex conjugate of $\uB$, and $G(\hat z,z)=1/(2\pi)(\hat z- z)^2$ is the Greens
function of a dipole. We thus have to integrate over all sources at points $z=x+iy$ in
the shaded area $\Omega$ in Figure~\ref{fig:geo} and weigh their contribution with the
Greens function $G(\hat z,z)$ to obtain the field $\uB$ at the point $\hat z$. 
\par
\begin{figure}[tb]
\begin{center}
\includegraphics[width=0.47\textwidth]{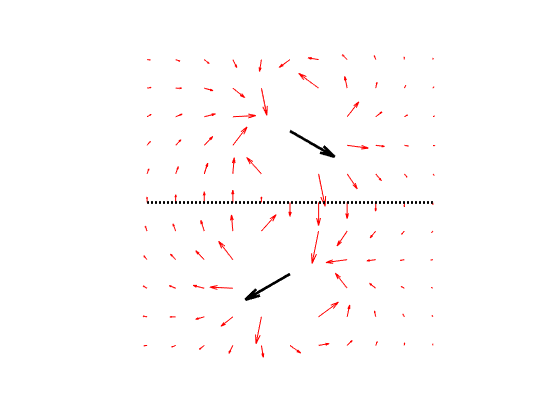}
\includegraphics[width=0.47\textwidth]{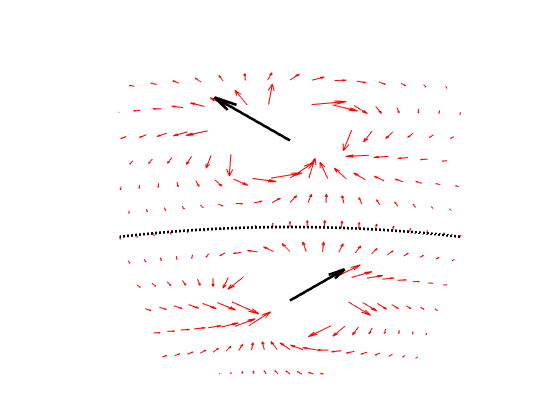}
\end{center}
\caption{\label{fig:im}Dipole images of a plane a cylindrical surface.} 
\end{figure}

The magnetic fields on a boundary, assumed to have infinite permeability, are purely
normal, because the field lines prefer go through the high-permeability material rather
than through air. We first consider a single dipole close to and below a planar magnetic
boundary, as shown on the left-hand side in Figure~\ref{fig:im}. The orientation of the
image above the plane has the same normal component of the dipole moment below the
plane, but the sign of the tangential component is reversed. This is easy to see when
considering a magnetic dipole that is composed of two very close monopoles; the one closer
to the boundary also must have its image closer to the boundary. We illustrate the field of
this configuration by red arrows visible on the left-hand side in Figure~\ref{fig:im}.
We thus find that for a boundary along the $x$-direction, the image dipole must be the
negative and complex conjugate $\tilde{\uB}_r=-\uB_r^{\ast}$ of the original dipole $\uB_r$
and it must be located at the mirrored position on the ``other'' side of the boundary.
\par
Let us now consider a cylindrical boundary with radius $R$ and centered at the
origin. Inside, there is a single dipole source located at radius $r$ and placed
on the vertical axis. Later we will remove this restriction. We point out that
sources, inside the cylinder, have images that are located outside the
cylinder at a distance $R^2/r$ from its center~\cite{APB}. This is also true
for two infinitesimally close (virtual) magnetic monopoles that make up a dipole;
their distance increases by a factor $(R/r)^2$. For very close
points the sagitta is almost equal to the length of the arc between points and
that increases by $R/r$ when going from $r$ to the cylinder at $R$. It then
increases by $(R^2/r)/R=R/r$ once again when going from the cylinder to the
image point. Therefore the strength the image dipole must be larger by a
factor $(R/r)^2$ compared to the original dipole, which is consistent with
the reasoning from~\cite{PLONESY}. The right-hand side in Figure~\ref{fig:im}
shows the original dipole below and the image above the cylindrical boundary
as well as the field they create, shown by the red arrows. We see that the image
is stronger and further away from the boundary causing the field to become
purely tangential close to the boundary. 
\par
\begin{figure}[tb]
\begin{center}
\includegraphics[width=0.5\textwidth]{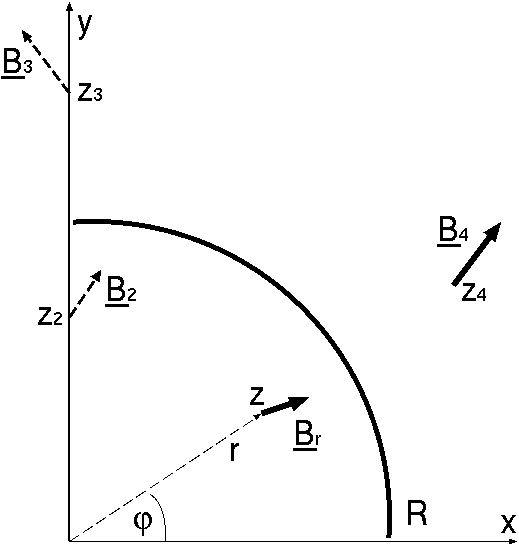}
\end{center}
\caption{\label{fig:cim}The image dipole at $z_4$ is constructed by first rotating
  the original dipole $\uB_r$ from $z$ to the imaginary axis at $z_2$, followed by
  scaling its magnitude with $(R/r)^2$, complex conjugating it, and moving it to $z_3$.
  Finally the dipole is rotated back and arrives at $z_4$. See the text for further
  explanations.}
\end{figure}
We now generalize the geometry to account for arbitrary positions of the original
dipole inside the cylinder at $z=x+iy=r e^{i\vp}$ such as the dipole labeled $\uB_r$
in Figure~\ref{fig:cim}. In order to determine its image, we first rotate it by the
angle $\pi/2-\vp$ to lie on the imaginary axis at $z_2$, where it becomes
$\uB_2=e^{i(\pi/2-\vp)}\uB_r$. Applying the procedure from the previous paragraph
we obtain its image
\begin{equation}
  \uB_3=-\frac{R^2}{r^2}\uB_2^{\ast}=-\frac{R^2}{r^2}e^{-i(\pi/2-\vp)} \uB_r^{\ast}
\end{equation}
at $z_3$. Rotating back with angle $-(\pi/2-\vp)$ we finally obtain the
image $\uB_4$, located at $z_4=R^2/z^{\ast}$
\begin{equation}\label{eq:imdip}
  \uB_4=-e^{-i(\pi/2-\vp)}\frac{R^2}{r^2}e^{-i(\pi/2-\vp)}  \uB_r^{\ast}
  =-\frac{R^2 e^{-i\pi}}{r^2e^{-2i\vp}}  \uB_r^{\ast} = \frac{R^2}{\left(z^{\ast}\right)^2} \uB_r^{\ast}\ ,
\end{equation}
where $z^{\ast}=x-iy$ is the complex conjugate of the original dipole's location.
Finding the image thus requires scaling with $(R/z^{\ast})^2$ and taking the complex
conjugate of the original dipole $\uB_r$.
\section{Shielded continuously rotating multipole ring}
\label{sec:crot}
%
\begin{figure}[tb]
\begin{center}
\includegraphics[width=0.9\textwidth]{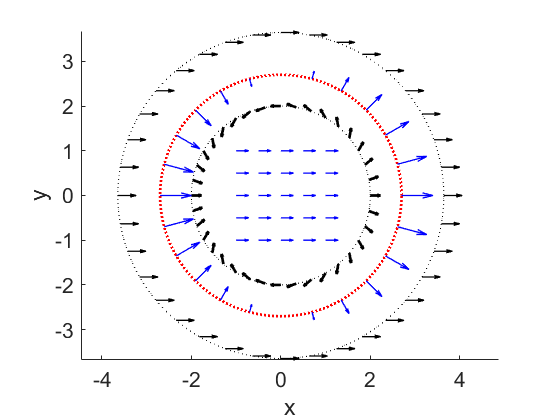}
\end{center}
\caption{\label{fig:cr}The original dipoles are visualized by the arrows on the
  inner dotted black circle. The rotate twice as one goes along the circle,
  consistent with a tumbling factor of $k = 2$. The shielding is shown as the red
  circle whereas the image dipoles, calculated from Equation~\ref{eq:imdip}
  are located further outside and are also given by black arrows. The blue
  arrows on the shielding is calculated by adding the field from all dipoles,
  weighted by the Green's function $G(\hat z,z)$, on the shielding.}
\end{figure}
We now proceed to calculate the field and its multipole contents at the center of
the cylinder created by the the permanent-magnet material and its images. Adding the
images caused by  a cylindrical enclosure to Equation~\ref{eq:Bdip} we obtain
\begin{equation}\label{eq:crmr}
  \uB^{\ast}(\hat z)=\frac{1}{2\pi}\int_{\Omega} \left[
    \frac{\uB_r}{(\hat z- z)^2} + \frac{R^2}{z^{\ast2}}\frac{\uB_r^{\ast}}{(\hat z- R^2/z^{\ast})^2}\right] dx dy\ .
\end{equation}
Using the identity
\begin{equation}
\frac{1}{(\hat z-z)^2} = \frac{d}{dz}\left(\frac{1}{\hat z-z}\right) 
= -\frac{d}{dz} \left(\sum_{m=0}^{\infty}\frac{\hat z^{m}}{z^{m+1}}\right)
= \sum_{m=0}^{\infty} \frac{(m+1) \hat z^{m}}{z^{m+2}}
\end{equation}
we rewrite Equation~\ref{eq:crmr} as a multipole expansion around the origin
\begin{eqnarray}\label{eq:uB}
  \uB^{\ast}(\hat z)
  &=&\frac{1}{2\pi}\int_{\Omega}dx dy\sum_{m=0}^{\infty}\left[
      \frac{(m+1) \uB_r}{z^{m+2}} + \frac{R^2}{z^{\ast2}}  \frac{(m+1)\uB_r^{\ast}}{(R^2/z^{\ast})^{m+2}} \right] \hat z^m
  \nonumber\\
  &=& \sum_{m=0}^{\infty}\left\{ \frac{m+1}{2\pi}\int_{\Omega}dx dy
      \left[ \frac{\uB_r}{z^{m+2}} + \frac{\uB_r^{\ast} }{R^{2m+2}}(z^{\ast})^m\right] \right\} \hat z^m\ .
\end{eqnarray}
Introducing cylindrical coordinates with $z=x+iy=re^{i\vp}$ and $dx dy=r dr  d\vp$ we
arrive at
\begin{equation}\label{eq:Bphi}
  \uB^{\ast}(\hat z)=\sum_{m=0}^{\infty}\frac{m+1}{2\pi}\int_{r_i}^{r_o} dr \int_0^{2\pi}d\vp
    \left[\frac{\uB_r}{r^{m+1}} e^{-i(m+2)\vp} + \frac{\uB_r^{\ast}}{R^{2m+2}} r^{m+1} e^{-im\vp}\right]\hat z^m
\end{equation}
where we assume that the magnetic material extends from inner radius $r_i$ to outer
radius $r_o$.
\par
The multipolarity is determined by the tumbling factor $k$ that specifies how often the
easy axis of the permanent magnet material rotates as a function of the azimuthal
angle $\vp.$ For example, dipoles have $k=2$ and quadrupoles have $k=3$. The easy
axis is thus given by $\uB_r=B_re^{ik\vp}$. Here $B_r=\vert B_r\vert e^{i\psi}$ describes
the magnitude and direction of the permanent-magnet material at the positive horizontal
axis at $\vp=0$. As an aside note that in Figure~\ref{fig:cr} the easy axis points
along the horizontal axis, which makes $\psi=0$. Without this restriction
Equation~\ref{eq:Bphi} then leads to
\begin{eqnarray}
  \uB^{\ast}(\hat z)&=&|B_r|\sum_{m=0}^{\infty}\frac{m+1}{2\pi}\int_{r_i}^{r_o} dr \int_0^{2\pi}d\vp\\
  &&\qquad\times
  \left[e^{i\psi}\frac{1}{r^{m+1}}e^{i(k-m-2)\vp} + e^{-i\psi}\frac{r^{m+1}}{R^{m+2}}e^{-i(k+m)\vp}\right]\hat z^m\ ,\nonumber
\end{eqnarray}
where the first term in the square brackets denotes the field created by the original dipoles
already derived in~\cite{HALBACH} and the second term describes the contribution from the
image fields. Remarkably, due to the high symmetry of the problem, this second term always
vanishes, because integrating $e^{-i(k+m)\vp}$ between zero and $2\pi$ vanishes for
tumbling factors with $k>0$.
\par
In order to analyze this surprising observation, we consider the simple magnet with
continuously rotating
dipoles and $k=2$ in Figure~\ref{fig:cr}, where the original dipoles are placed on the inner
ring denoted by black dots. Note how the dipoles rotate twice along the ring. The shielding
is shown as the red dotted line slightly outside of the original dipoles. The image dipoles,
calculated from Equation~\ref{eq:imdip}, are shown on the outer dotted ring, where, for $k=2$
they always point to the right. We verify that sum of the original and the image fields is
always normal on the shielding, which is shown by the superimposed blue arrows. We also
calculate the magnetic field inside the cylindrical shielding, which is shown by the
right-pointing arrows near the center. Moreover, inspecting the numerical values shows that
the contribution from the images to the field on the inside is negligible.
\par
We conclude that for a continuously varying dipole field shielding the magnet will not affect
the magnetic field on the inside. But what about segmented magnets?
\section{Shielded and segmented multipoles}
\label{sec:seg}
\begin{figure}[tb]
\begin{center}
\includegraphics[width=0.6\textwidth]{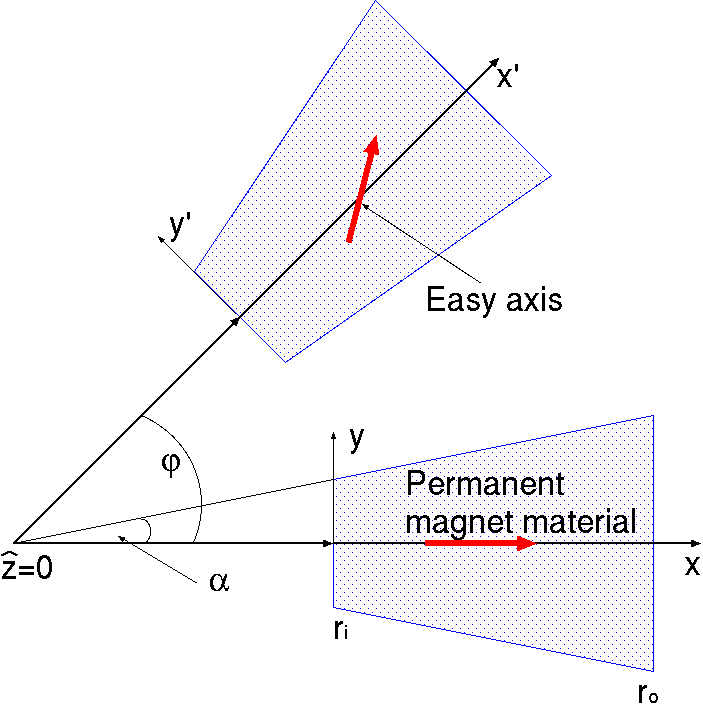}
\end{center}
\caption{\label{fig:pmt}Two segments of a segmented multipole that extends radially
  from $r_i$ to $r_o$. Each segment has an azimuthal width, given by $2\alpha$. Note
  the orientation of the easy axis, determined by the tumbling factor $k$, changes
  from segment to segment.}
\end{figure}
%
Figure~\ref{fig:pmt} shows two of the segments that are part of a full multipole.
We start by calculating the contribution of the field $\uB^{\ast}$ of the segment
that is lying on the horizontal axis to the multipoles around $\hat z=0$.
Rewriting Equation~\ref{eq:uB} leads to
\begin{equation}\label{eq:uBs}
  {\uB}^{\ast}(\hat z) = \sum_{m=0}^{\infty}\left\{
    \frac{m+1}{2\pi}\left[ {\uB}_r I_m+ {\uB}_r^{\ast}J_m\right]\right\}\hat z^m
\end{equation}
where ${\uB}_r$ describes the easy axis of the segment on the horizontal axis. The
integrals $I_m$ and $J_m$ are defined by
\begin{equation}\label{eq:II}
  I_m=\int_{\Omega}\frac{dx dy}{z^{m+2}}
  \qquad\mathrm{and}\qquad
  J_m=\frac{1}{R^{2m+2}}\int_{\Omega} \left(z^{\ast}\right)^m dx dy\ ,
\end{equation}
where $I_m$ describes the contribution of the ``real'' permanent magnets and $J_m$
that of their images.
We calculate these integrals by suitably parameterizing the trapezoidal region $\Omega$
using the notation introduced in Figure~\ref{fig:pmt}, which gives us
\begin{eqnarray}\label{eq:Ione}
  I_m&=&\int_{r_i}^{r_o}dx\int_{-x\tan\alpha}^{x\tan\alpha}\frac{dy}{(x+iy)^{m+2}}\nonumber\\
  &=& \frac{-i}{m+1}\int_{r_i}^{r_o}\frac{dx}{x^{m+1}}
      \left[\frac{1}{(1+i\tan\alpha)^{m+1}}-\frac{1}{(1-i\tan\alpha)^{m+1}}\right]\\
     &=&\frac{2}{m+1}(\cos\alpha)^{m+1}\sin((m+1)\alpha)G_m(r_i,r_o)\ ,\nonumber
\end{eqnarray}
where we introduce the abbreviation $G_m(r_i,r_o)$ by 
\begin{equation}\label{eq:GG}
  G_m(r_i,r_o)=\int_{r_i}^{r_o}\frac{dx}{x^{m+1}}=
  \left\{\begin{array}{ll}
      \log(r_o/r_i) & \mathrm{for}\ m=0\\
      \frac{1}{m}\left(\frac{1}{r_i^m}-\frac{1}{r_o^m}\right) & \mathrm{otherwise.}
  \end{array}\right.
\end{equation}
For the contribution of the image fields we evaluate $J_m$ and obtain
\begin{eqnarray}\label{eq:Itwo}
  J_m&=& \frac{1}{R^{2m+2}}\int_{r_i}^{r_o}dx\int_{-x\tan\alpha}^{x\tan\alpha} (x-iy)^m dy \nonumber\\
     &=&\frac{i}{R^{2m+2}(m+1)}\left[(1-i\tan\alpha)^{m+1} - (1+i\tan\alpha)^{m+1}\right]  \int_{r_i}^{r_o} x^{m+1}dx
  \nonumber\\
     &=&\frac{2\sin((m+1)\alpha)}{(m+1)(m+2)(\cos\alpha)^{m+1}}H_m(r_i,r_o,R)
\end{eqnarray}
with
\begin{equation}\label{eq:HH}
 H_m(r_i,r_o,R)=\frac{1}{R^{2m+2}}\int_{r_i}^{r_o} x^{m+1}dx=\frac{r_o^{m+2}-r_i^{m+2}}{R^{2m+2}}\ .
\end{equation}
Here we have to keep in mind that for a magnet composed of $M$ segments $2\alpha=2\pi/M$.
After inserting $I_m$ and $J_m$ in Equation~\ref{eq:uBs} we finally obtain
\begin{equation}
  \uB^{\ast}(\hat z) = \sum_{m=0}^{\infty} \left\{ {\uB}_r g_m(\alpha) G_m( r_i,r_o)
    + {\uB}_r^{\ast}h_m(\alpha) H_m(r_i,r_o,R) \right\} \hat z^m
\end{equation}
with
\begin{eqnarray}\label{eq:gh}
  g_m(\alpha) &=&\frac{1}{\pi}(\cos\alpha)^{m+1}\sin((m+1)\alpha)\nonumber \\
  h_m(\alpha) &=&\frac{\sin((m+1)\alpha)}{\pi(m+2)(\cos\alpha)^{m+1}}
\end{eqnarray}
for the multipole coefficients generated by a single segment.
\par
\begin{figure}[tb]
\begin{center}
\includegraphics[width=0.8\textwidth]{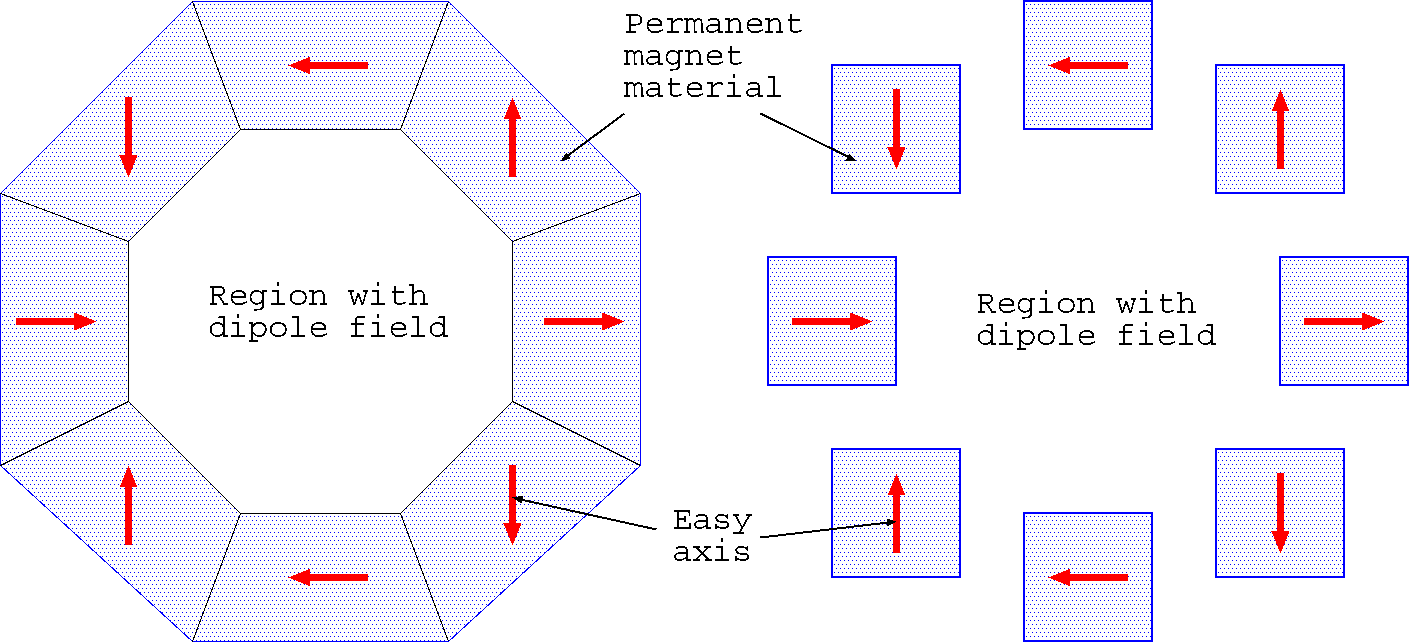}
\end{center}
\caption{\label{fig:dip}Segmented dipole (left) and a dipole made of permanent-magnet cubes (right).}
\end{figure}
In the next step we have to add the contributions of the $M$ segments by noting that
each segment $j$ with $0\le j\le M-1$ is rotated by an angle $\vp=2\pi j/M$ with respect
to the one lying on the horizontal axis. This changes the integration variable from
$z$ to $z'=x'+iy'=e^{i\vp}(x+iy)=e^{i\vp}z$. Moreover, the angle of the easy axis changes
by $k\vp$. Thus we can add up contributions from the segments by ornamenting the
easy axis $\uB_r$ and $\uB_r^{\ast}$ in Equation~\ref{eq:uBs} and the integrals $I_m$ and
$J_m$ from Equations~\ref{eq:Ione} and~\ref{eq:Itwo} with appropriate phase factors to
obtain for the field due to all segments
\begin{eqnarray}
  \uB^{\ast}(\hat z)
  &=&\sum_{m=0}^{\infty}\left\{
    {\uB}_r g_m(\alpha)G( r_i,r_o)\sum_{j=0}^{M-1}e^{2\pi i(k-m-2)j/M}\right. \nonumber\\
  &&\qquad\left.  +{\uB}_r^{\ast}h_m(\alpha) H(r_i,r_o,R)\sum_{j=0}^{M-1}e^{2\pi i(m-k)j/M}\right\} \hat z^m\ .
\end{eqnarray}
The first sum with the exponential factors vanishes unless the $k-m-2$ is a multiple of $M$,
or $m=k-2+\nu M$ for some integer $\nu$ in which case the sum evaluates to $M$. Likewise,
the second sum vanishes unless $m-k$ is a multiple of $M$, or $m=k+\nu M$. This allows us
to write the field from all segments as
\begin{eqnarray}\label{eq:Bi}
  \uB^{\ast}(\hat z)
  &=&M\sum_{m=0}^{\infty}\sum_{\nu=0}^{\infty}\left\{{\uB}_rg_m(\alpha)G_m(r_i,r_o)\delta_{m,k-2+\nu M}\right.\nonumber\\
  &&\left.\qquad\qquad + {\uB}_r^{\ast}h_m(\alpha) H_m(r_i,r_o,R)\delta_{m,k+\nu M}\right\}\hat z^m\ ,
\end{eqnarray}
where $g_m(\alpha)$ and $h_m(\alpha)$ are defined in Equation~\ref{eq:gh},
$G_m(r_i,r_o)$ in Equation~\ref{eq:GG}, and $H_m(r_i,r_o,R)$ in Equation~\ref{eq:HH}.
\par
As an illustration we consider a dipole magnet ($k=2$) made of $M=8$ segments, whose
inner radius $r_i$ is 10\,mm and whose outer radius $r_o$ is 20\,mm. We assume that the
shielding cylinder has a radius of $R=22\,$mm. The main contribution comes from the first
term in Equation~\ref{eq:Bi} for $m=0$, in which case $g_m(\alpha)=0.115$
with $\alpha=\pi/M$ and
$G_m(r_i,r_o)=\log(2)$. The first term --- the dipole --- is constant and has magnitude $0.62\uB_r$.
The smallest non-zero second term is for $m=k=2$, which describes a sextupole. Its magnitude
at a radius $|\hat z|=r_i/2$ becomes $M h_2(\alpha)H(r_i,r_o,R){\uB}_r^{\ast} (r_i/2)^2
\approx 0.025{\uB}_r^{\ast}$ or about 4\,\% of the dipole component.
\par
For quadrupole magnets ($k=3$) made of $M=8$ magnets and otherwise the same geometry as the
dipole from the previous paragraph, we find that the direct field from the ``real'' permanent
magnets at radius $|\hat z|=r_i/2$ has magnitude $0.38\uB_r$ whereas the first contribution
(octupolar) of the image fields is $0.005\uB_r^{\ast}$, or a little over 1\,\% of the quadrupole
component.
\section{Shielded multipoles made of permanent-magnet cubes}
\label{sec:cube}
%
\begin{figure}[tb]
\begin{center}
\includegraphics[width=0.6\textwidth]{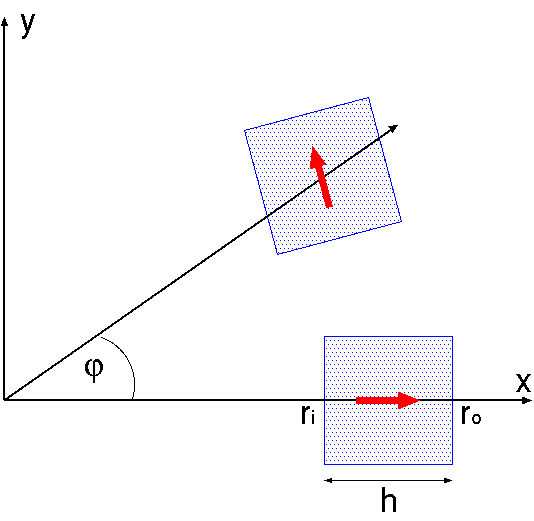}
\end{center}
\caption{\label{fig:cube}Geometry with variables used to calculate the fields
  of a permanent magnet multipole composed of cubes.}
\end{figure}
Permanent-magnet cubes are easier to find and less expensive than
the trapezoidal segments used in Section~\ref{sec:seg}. We therefore analyze how
shielding affects multipoles that are constructed of cubes~\cite{APB,STRAP}. The
right-hand side in
Figure~\ref{fig:dip} illustrates such a magnet that generates a purely horizontal
magnetic field. We calculate the field from Equations~\ref{eq:uBs} and~\ref{eq:II}
after adapting the integration region $\Omega$ to reflect the cubic shape shown in
Figure~\ref{fig:cube}. The first integral $I_m$ then becomes
\begin{eqnarray}\label{eq:Imcube}
  I_m&=&\int_{r_i}^{r_o}dx\int_{-h/2}^{h/2}\frac{dy}{(x+iy)^{m+2}}\nonumber\\
     &=&\frac{i}{m+1}\int_{r_i}^{r_o} \left[\frac{1}{(x+ih/2)^{m+1}}-\frac{1}{(x-ih/2)^{m+1}}\right] dx\ .
\end{eqnarray}
For $m=0$ the integral leads to
\begin{eqnarray}\label{eq:Icube0}
  I_0&=&h\int_{r_i}^{r_o}\frac{dx}{x^2+h^2/4} = 2\arctan(2r_o/h)-2\arctan(2r_i/h)\nonumber\\
     &=& 2\arctan(h^2/2\bar r^2)\ ,
\end{eqnarray}
where we introduce $\bar r=(r_o+r_i)/2$ and use the identity $\arctan x -\arctan y
=\arctan((x-y)/(1+xy))$ in order  to simplify the last equality. For $m=1$ the integral
becomes
\begin{equation}
  I_1=\frac{-h}{2}\left[\frac{1}{r_o^2+h^2/4}-\frac{1}{r_i^2+h^2/4}\right]
  =\frac{h^2\bar r}{\bar r^4+h^4/4}\ .
\end{equation}
For general $m>0$ we perform the integral over $x$ and then introduce the abbreviations
\begin{equation}\label{eq:rhobeta}
  \begin{array}{ll}
  \rho_o^2=r_o^2+h^2/4 \qquad &\rho_i^2=r_i^2+h^2/4\\
    \beta_o=\arctan(h/2r_o)\qquad  &\beta_i=\arctan(h/2r_i)\ .
  \end{array}
\end{equation}
This allows us to express $r_o\pm ih/2=\rho_oe^{\pm i \beta_o}$ and
$r_i\pm ih/2=\rho_ie^{\pm i \beta_i}$ and turn Equation~\ref{eq:Imcube} into
\begin{eqnarray}\label{eq:IImcube}
I_m
&=& \frac{-i}{m(m+1)}\left[\frac{1}{(r_o+ih/2)^m} -\frac{1}{(r_i+ih/2)^m}
  -\frac{1}{(r_o-ih/2)^m}+\frac{1}{(r_i-ih/2)^m}\right]\nonumber\\
  &=& \frac{-i}{m(m+1)}\left[\frac{e^{-im\beta_o}-e^{im\beta_o}}{\rho_o^m}
      - \frac{e^{-im\beta_i}-e^{im\beta_i}}{\rho_i^m}\right]\nonumber\\
  &=& \frac{2}{m(m+1)}\left[\frac{\sin(m\beta_i)}{\rho_i^m}-\frac{\sin(m\beta_o)}{\rho_o^m}\right]\ .
\end{eqnarray}
For cubic magnets the integral $J_m$ corresponding to Equation~\ref{eq:Itwo} becomes  
\begin{eqnarray}\label{eq:JJmcube}
  J_m&=& \frac{1}{R^{2m+2}}\int_{r_i}^{r_o}dx\int_{-h/2}^{h/2} (x-iy)^m dy \nonumber\\
  &=& \frac{-i}{(m+1)R^{2m+2}}\int_{r_i}^{r_o}\left[(x+ih/2)^{m+1}-(x-ih/2)^{m+1}\right] dx\\
  &=& \frac{2}{(m+1)(m+2)}\left[\frac{\rho_o^{m+2}}{R^{2m+2}}\sin((m+2)\beta_o)-\frac{\rho_i^{m+2}}{R^{2m+2}}\sin((m+2)\beta_i)\right]\ ,
      \nonumber
\end{eqnarray}
where $\rho_o$, $\rho_i$, $\beta_o$, and $\beta_i$ are defined in Equation~\ref{eq:rhobeta}.
\par
Adding the fields from $M$ cubes with their appropriate rotations progresses in much the
same way as in the previous section. The easy axis of $\uB_r$ rotates by $e^{2\pi i k j/M}$ and
the powers $z$ by $e^{2\pi i j/M}$ for cube number $j$ with $0\le j\le M-1$. Again, the sum of
the $M$ permanent-magnet cubes becomes zero except for multipoles $m=k-2+\nu M$ and for
multipoles $m=k+\nu  M$ for the image fields. Only the numerical factors differ from those
in Section~\ref{sec:seg}; for the cubes the field thus becomes
\begin{equation}\label{eq:Bc}
  \uB^{\ast}(\hat z)
  =M\sum_{m=0}^{\infty}\sum_{\nu=0}^{\infty}\frac{m+1}{2\pi}\left\{{\uB}_rI_m\delta_{m,k-2+\nu M}
     + {\uB}_r^{\ast}J_m\delta_{m,k+\nu M}\right\}\hat z^m\ ,
\end{equation}
with $I_m$ from Equation~\ref{eq:Icube0} or~\ref{eq:IImcube} and $J_m$ from Equation~\ref{eq:JJmcube}.
\par
For a dipole with tumbling factor $k=2$ the direct field from the cubes generates multipoles
$m=\nu M$ and the image fields generate multipoles $m=2+\nu M$, such that the two lowest-order
contributions are the dipole field $m=0$ from the cubes and $m=2$ from the images. The fields
are thus given by 
\begin{eqnarray}
  \uB^{\ast}(\hat z)
  &\approx&\frac{M}{2\pi}\left[{\uB}_r I_0 + {\uB}_r^{\ast} J_2 \hat z^2\right]\\
  &=&\frac{M}{\pi} \left[{\uB}_r\arctan(h^2/2\bar r^2) + {\uB}_r^{\ast}
\left(\frac{\rho_o^{4}}{4R^{6}}\sin(4\beta_o)-\frac{\rho_i^4}{4R^6}\sin(4\beta_i)\right)\hat z^2
      \right]\ .\nonumber
\end{eqnarray}
Notably, the image fields decrease with the sixth power of the radius $R$ of the
shielding. Increasing $R$ therefore efficiently helps to reduce the unwanted sextupole
component. We refrain from discussing numerical examples, because they are similar to
those of segmented magnets from Section~\ref{sec:seg}.
\section{Conclusions and outlook}
\label{sec:conc}
From expressions for the image dipoles outside a magnetically shielding cylinder,
we calculated the additional multipole components they cause in Halbach-type
multipoles. These additional fields are typically rather small, which can be already
expected from the perfect cancellation of the additional fields for a idealized
multipole with continuously rotating easy axis (Section~\ref{sec:crot}). But even
for segmented or cube-based magnets, the additional fields are small and are attenuated
at least by a factor $(\tilde r/R)^6$ for dipoles and $(\tilde r/R)^8$ for quadrupoles.
Here $\tilde r$ typically is the outer radius of the permanent magnet material.
Making the shielding cylinder even a little larger thus reduces the additional
fields substantially. The multipolarity $m$ of the additional fields is given by
$m=k+\nu M$ where $k=2$ for dipoles and $k=3$ for quadrupoles. $M$ is the number
of segments or cubes.
\par
In general, Equations~\ref{eq:Bi} and~\ref{eq:Bc} can be used to calculate the
multipolarity and the magnitude  of the additional fields. Overall, we find that
the shielding has a small influence on the field quality in Halbach-type multipoles.
\par
This work was produced in part by Jefferson Science Associates, LLC under
Contract No. AC05-06OR23177 with the U.S. Department of Energy. Publisher
acknowledges the U.S. Government license and provide public access under
the DOE Public Access Plan (\url{http://energy.gov/downloads/doe-public-access-plan}).
%
%
\bibliographystyle{plain}

\end{document}